\documentclass[runningheads]{llncs}
\usepackage{soul}
\usepackage{subfig}
\usepackage{graphicx}
\usepackage{booktabs}
\usepackage{url}
\usepackage{caption}

\begin{document}
\title{Introducing the Task-Aware Storage I/O (TASIO) Library}
%
%
\author{Aleix Roca Nonell\inst{1}\orcidID{0000-0002-6715-3605} \and
\mbox{Vicen\c{c} Beltran Querol\inst{1}\orcidID{0000-0002-3580-9630}} \and
\mbox{Sergi Mateo Bellido\inst{1}\orcidID{0000-0002-2169-2240}}}
\authorrunning{A. Roca et al.}
%
\institute{Barcelona Supercomputing Center, Spain
\email{\{aleix.rocanonell,vbeltran,sergi.mateo\}@bsc.es}\\
\url{https://www.bsc.es}}
\maketitle              
\begin{abstract}


Task-based programming models are excellent tools to parallelize and seamlessly load balance an application workload. However, the integration of I/O intensive applications and task-based programming models is lacking. Typically, I/O operations stall the requesting thread until the data is serviced by the backing device. Because the core where the thread was running becomes idle, it should be possible to overlap the data query operation with either computation workloads or even more I/O operations. Nonetheless, overlapping I/O tasks with other tasks entails an extra degree of complexity currently not managed by programming models' runtimes. In this work, we focus on integrating storage I/O into the tasking model by introducing the Task-Aware Storage I/O (TASIO) library. We test TASIO extensively with a custom benchmark for a number of configurations and conclude that it is able to achieve speedups up to 2x depending on the workload, although it might lead to slowdowns if not used with the right settings.

\end{abstract}


\vfill
\begin{flushleft}
\noindent The final authenticated publication is available online at \linebreak
\hspace*{0.5em} \url{https://doi.org/10.1007/978-3-030-28596-8_19}
\end{flushleft}

\section{Introduction}
\label{sec:introduction}

In the road to exascale, it is essential to ensure the most efficient use of hardware resources. The increasing number of cores and hardware threads in modern computers requires an extra effort for application programmers to properly distribute work among cores. In particular, the penalization for not properly balancing an application workload is aggravated given that a single core in the application's critical path is able to keep all the other cores idle until its work finishes.

Programming models have proven to be a powerful tool to ease the processes of parallelizing applications regardless of their use case. Notably, task-based programming models are especially suitable to perform transparent load balancing by simply adjusting the size and/or number of tasks. However, because of its generality, programming models refrain from specifying use cases for I/O intensive applications.

I/O intensive applications have a huge impact on the system's resource usage. Typically, I/O operations require of operating system assistance to interface with a particular hardware device. Such devices are slower than the main processing units which force the thread issuing the I/O request to either continuously poll for the completion or to block inside the operating system. In the first case, power and time is wasted in not truly productive work. In the second case, the core becomes idle allowing other system threads to run on it. However, because I/O operations are usually more expensive than the system's threads computing requirements, most of the time the core will be idling anyways. Typically, this problem is solved by using asynchronous functions to avoid the thread blocking on the operation. However, the application's design complexity increases substantially when trying to combine asynchronous calls with task-based programming models effectively.

OmpSs-2 is a task-based programming model (see Section~\ref{sec:bkg_ompss}) whose runtime is "asynchronous-aware", which means that provides an interface to register tasks performing asynchronous operations. In order to make use of such interface, in this paper we present a new library named \texttt{TASIO} which replaces synchronous I/O system calls by their asynchronous counterparts and notifies the runtime to schedule other tasks on the core while the operation is being serviced.

Hence, in this article we present the following contributions: 1) We propose the TASIO library to enable the conversion of synchronous to asynchronous operations and integrate it with "asynchronous-aware" runtimes 2) We present a task-based synthetic benchmark which simulates interleaved computation and I/O workloads and 3) We explore the results space of the synthetic benchmark for a number of configurations and detail the conclusions learned.


\section{Background}
\label{sec:background}

\subsection{The OmpSs-2 programming model}
\label{sec:bkg_ompss}

OmpSs-2~\cite{bsc:ompss,bsc:ompss2} is a task-based programming model developed at the Barcelona Supercomputing Center (BSC) with the objective of early-testing novel features for the tasking model that might influence the development of the OpenMP programming model~\cite{bsc:openmp1,bsc:openmp2}. The main focus of OmpSs-2 is in both asynchronous parallelism and device heterogeneity (distribute work among different devices such as systems' cores, GPUs, and FPGAs). The source to source Mercurium compiler and the Nanos6 runtime are the BSC's implementation of the OmpSs-2 model. Mercurium translates source code pragmas into Nanos6 library calls while Nanos6 manages the application's execution flow at runtime.

In a task-based programming model, all units of parallelism are expressed as tasks. A task is an enclosed sequence of instructions specified by the developer that must be executed sequentially. Multiple tasks can be executed in parallel as long as all their dependencies have been fulfilled. Dependencies express which data is required by tasks to perform its computation and which data it produces. Dependencies are expressed simply by specifying which variables a task uses as input, output or both. The actual execution sequence of tasks is determined by the runtime.

\subsection{Linux Kernel Asynchronous I/O interfaces}
\label{bkg:aio}

Synchronous I/O requests typically block\footnote{The operation might return immediately, i.e. not block, if the system page cache already holds the requested data in the case of reads or if the page cache has enough free space as to defer the operation for a later time in the case of writes.} while the request is being processed. Instead, asynchronous I/O requests are intended to avoid blocking by separating the operation into two parts: submission of the request and check for completion. There are two standard implementations for asynchronous I/O in modern Linux machines: The Linux Kernel native AIO and the POSIX AIO.



The Linux Kernel native asynchronous I/O interface\cite{aio} consists in a set of system calls to submit and monitor I/O requests independently from the set of typical synchronous system calls. AIO requests are submitted in a context~\footnote{A context is basically a queue of requests} created beforehand. When the submission operation returns, it is possible to check for the request status and to explicitly block until any of the requests in the context have finished. The POSIX AIO, instead, simulates Kernel AIO support by simply delegating synchronous I/O operations to a pool of threads.

Similarly to the synchronous approach, it is likely that the submission of an asynchronous request is completed just after it is submitted because of the effects of the page cache. For this reason, Linux AIO is only useful when the page cache is bypassed (non-buffered I/O). However, there are other system specific factors that might prevent the AIO requests to actually be asynchronous such as filesystem limitations. For instance, the ext4 filesystem mandates that the AIO operation should not modify the file metadata~\cite{wiki:ext4aio} such as when enlarging a file due to a write operation.

\section{Related Work}
\label{sec:related-work}

Scientific application codes have historically used custom thread implementations to manage parallelism. Because these applications are usually complex and moving from the classic thread paradigm to a task-based solution is not usually simple, most of them refrain from changing their parallel scheme. Also, storage I/O has been traditionally implemented in sequential bursts due to constraints associated with legacy hard disks. Moreover, asynchronous I/O usually imposes strict constraints that are not always easy to meet. In consequence, there is not much literature focusing on the interaction of task-based programming models and asynchronous parallel I/O at the node level.

However, the Message Passing Interface (MPI) library is widely adopted and previous research exists on overlapping MPI communications with computation in task-based programming models. The Task-Aware MPI library (TAMPI)~\cite{bsc:tampi} tackles the problem of overlapping network communications with other workloads. It uses the OmpSs-2 pause/resume, external events and polling services APIs (see Section~\ref{sec:computio}) to minimize the time cores are idling while communicating over network. The TASIO library presented in this paper is highly inspired by TAMPI. 


\section{Computation and I/O Overlapping with OpenMP and OmpSs-2}
\label{sec:computio}

I/O operations usually rely on blocking system calls that stall the execution of work in cores. On task-based programming models, this entails a performance degradation because runtimes are not aware of when cores became idle and hence, are not able to run other tasks on them while I/O is being serviced. A common technique to overlap I/O and computation is to run asynchronous I/O operations instead of blocking ones. Yet the integration of asynchronous calls with task-based programming models is usually tedious. Task-based programming models work on the abstraction of data dependencies and execution flow. Data consumed or generated asynchronously needs to be tracked by the dependency system which means that a task must generate or consume the data. Nonetheless, asynchronous operations need to be checked for completion by either polling or callback, but neither of them are trivial to wrap within a task.

This section explores the proposals of both the OpenMP and OmpSs-2 programming models and it introduces the TASIO library based on the OmpSs-2 APIs.

\subsection{OpenMP}

The OpenMP 5.0 specification~\cite{spec:openmp} introduces the \texttt{detach} clause to the task construct with the purpose of delaying the completion of a task (possibly) long after its body has been executed. To complete a task with a detach clause it is necessary to, on the one hand, run it and, on the other hand, mark its event object (provided within the detach clause) as completed using an OpenMP API call. A task submitting an asynchronous operation will define the consumed or generated data in its dependencies and those will only be released when the task finishes. However, the task will not be completed until the code responsible to check for the asynchronous completion operation marks the associated detach event as satisfied.

The OpenMP specification gives complete freedom to the developer to decide how and when the completion checking code should be run. When working with frameworks providing callback support such as CUDA, running the OpenMP detach completion function is simple. However, when polling or blocking is needed the user needs to create its own thread and care must be taken to prevent this thread from overlapping with the computation of the other OpenMP threads.

\subsection{OmpSs-2}
\label{sec:ci-ompss2}

OmpSs-2 features two APIs~\footnote{The low level details of such APIs can be consulted in~\cite{bsc:tampi}} to deal with blocking operations: 

The \textit{pause/resume API} allows to pause the execution of the current task and to resume it later on. Pausing deschedules the current task and returns control back to the runtime which is able to run other tasks in the core where the first task was running. Once a task is paused, the next task could be run either on the same thread as the previous task or on a new thread. In the first case, the stack of the paused stack becomes buried below the stack of the next task and, hence it cannot be resumed until the second task finishes. This could lead to a deadlock for tasks that use two-sided messages APIs such as MPI, therefore, the Nanos6 runtime implements the second approach because of its genericity.

The \textit{external events API} does not stop the execution of a task, but it simply delays the release of its dependencies until all registered events have been fulfilled (similarly to the OpenMP detach clause). In other words, during the execution of a task, a number of events can be registered within Nanos6. Even if the task code finishes, it will not unblock the tasks that depend on this task until all events are satisfied. In consequence, asynchronously requested data cannot be used inside the same task that requests it, because its fulfillment is likely to occur after the task body is finished.

Both APIs rely on the OmpSs-2 \textit{polling services API} to periodically run a user-registered function within Nanos6. This function is run at strategic points to avoid disturbing other tasks. A possible use case for this functions is to poll for completion of registered asynchronous events. The exact method to check for completion depends on the kind of submitted asynchronous operation and is the (library developer) user responsibility to code. Once a completion event is detected, the polling function must either resume a task (in case using the pause/resume API) or decrement the event counter (for the external events API).


\subsection{The Task-Aware Storage I/O (TASIO) Library}
\begin{figure}
\centering
\includegraphics[width=10cm]{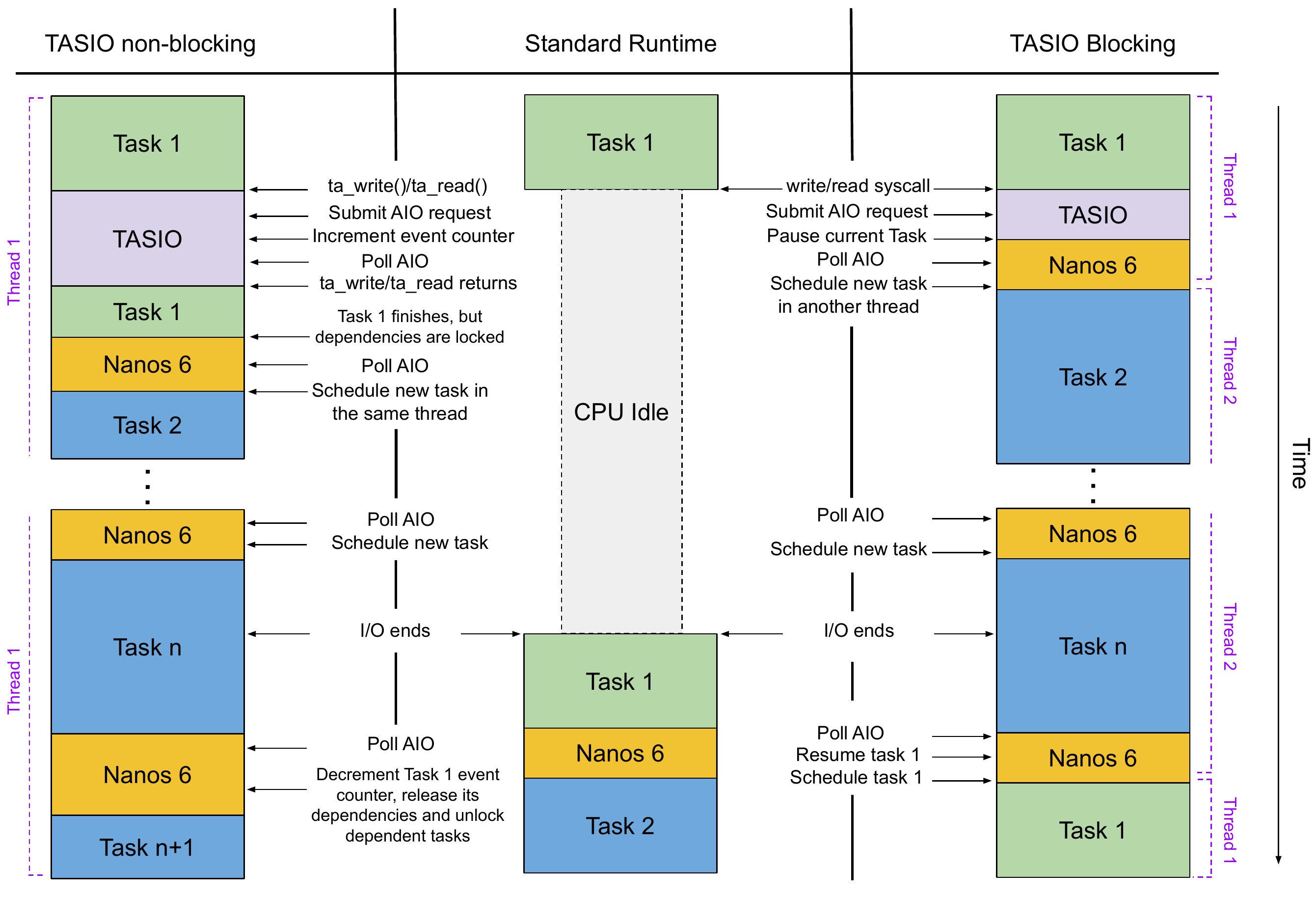}
\caption{TASIO Runtime execution flow example on a single core system.}
\label{fig:tasio_scheme}
\end{figure}

The TASIO library is similar to the TAMPI library; it provides both blocking and non-blocking APIs through OmpSs-2. The basic functionality is shown in figure \ref{fig:tasio_scheme}.

The TASIO blocking API (which uses the OmpSs-2 pause/resume API) defines wrappers for the \texttt{pread()}, \texttt{pwrite()}, \texttt{preadv()} and \texttt{pwritev()} syscalls (all Linux Kernel native AIO supported syscalls) which transparently call the asynchronous version of the intercepted syscall instead of the original one when applications are linked against it. After TASIO submits an AIO request it checks whether it has immediately completed or not and, if it is the case, the wrapper returns immediately as well. Otherwise, it sets the current task to the list of blocked tasks and transfers control to the runtime. The runtime is then able to execute other tasks in the current core while the I/O operation is being resolved. In this work, we focus on studying the pause/resume OmpSs-2 implementation that relies on creating extra threads on task pause because as explained in Section~\ref{sec:ci-ompss2} it is the most generic one. However, it is worth noting that the storage I/O does not suffer from the network I/O constraints and, therefore, would also work on the extra-thread-free version.

The TASIO non-blocking API (which uses the OmpSs-2 external events API) defines its own \texttt{ta\_pread()}, \texttt{ta\_pwrite()}, \texttt{ta\_preadv()} and \texttt{ta\_pwritev()} that behave as the pause/resume variant but instead of blocking the current thread it increments a task event counter and return immediately after submitting the AIO request.

At startup, TASIO registers a polling function within the Nanos6 runtime through the polling services API. Once a previously submitted asynchronous I/O request is completed, the polling function will retrieve it and either unlock the associated task if submitted through the pause/resume API or decrement its event counter for the external events API. The maximum amount of AIO petitions that TASIO can withstand at the same time is configured at 1000 by default. If at some point the maximum number of requests is reached, the offending request sleeps for 1ms and tries again\footnote{Smarter techniques could be used, but because this is a corner case we have simplified it for now.}.

It is worth noting that because TASIO uses the Linux Kernel AIO interface, all submitted I/O operations must be non-buffered and comply with the O\_DIRECT constraints as explained in Section~\ref{bkg:aio}.

Finally, we would like to point out that both TASIO and TAMPI can be combined in the same application and, in fact, this approach could be extended to other blocking mechanisms as long as asynchronous submission and non-blocking polling for completion mechanisms are supplied.

\subsection{OpenMP and TASIO/TAMPI support}

OpenMP support could be added to both the TASIO and TAMPI libraries (TAL hereinafter) non-blocking API by relying on the detach clause but it would not be possible to implement the TAL blocking API without a compatible pause/resume API. Nonetheless, there are three implementation limitations that would affect the non-blocking API performance and/or slightly increase its complexity. 

First and foremost, an additional thread would have to be created and managed by TAL to poll for AIO completions. The user would need to configure the thread's polling rate or at least be aware of the default one. However, as explained in Section~\ref{sec:ci-ompss2}, a runtime-aware completion thread would be more efficient.

Second, the OpenMP detach object has boolean semantics. Therefore, keeping track of multiple AIO submissions within a task requires an external counter. Also, a mechanism would be needed to associate detach objects and counters, such as a private TAL hash table. Even though, it would be particularly complex to combine multiple I/O functions of different task-aware non-blocking libraries within the same task because counters would be private per library. Keeping track of the number of requests within the detach OpenMP object and release its dependencies when zero is reached would simplify this detail.

And third (a minor detail), the user would need to feed the task context (detach object) to TAL functions when needed. OpenMP does not currently provide any means to obtain such context through an API call and therefore, it is not possible to retrieve it within TAL. However, adding such an API call within OpenMP should not be complex and would simplify the interface.

\section{Experimentation}
\label{sec:experimentation}

The TASIO exploitable benefits are highly dependent on the test environment. More precisely, it depends on the storage device, the number of cores and the application's I/O pattern. To cover as many cases as possible, we decided to implement a synthetic benchmark which we used to perform a deep scan on a number of configurations.

\subsection{The Task I/O Meter Benchmark (TIOM)}
\label{sec:tiom}

The Task I/O Meter Benchmark (TIOM) is a simple OmpSs-2 application that interleaves computation and I/O operations wrapped in tasks. The number of tasks, I/O block size per task, computation time per task and I/O pattern is configurable. I/O operations are performed on a user file and computation work is simulated by busy waiting in a loop. There are four main modes of operation that simulate different application's I/O patterns. Each mode creates a configurable amount of "task series" that can be completely run in parallel to other series. A task series is a chain of tasks that are bound by their dependencies.

In the \texttt{mix} mode, each task performs both computation and I/O, in this order. The \texttt{1to1} mode is similar, but computation and I/O are separated in different tasks bound with dependencies. In the \texttt{fjio} (fork-join I/O) and \texttt{fjc} (fork-join computation) modes, computation and I/O are also performed in separated tasks but, in \texttt{fjio}, each computation task depends on four I/O tasks and each I/O task depends on a single computation task. The \texttt{fjc} mode is similar to \texttt{fjio} but interchanging I/O per computation tasks.

The \texttt{mix} and \texttt{1to1} modes define an interleaved sequence of computation and I/O operations. However, only \texttt{mix} actually enforces this sequence. Instead, the more fine-grained \texttt{1to1} might allow sustaining more I/O operations in flight if a core happens to run multiple I/O tasks of different series instead of consistently alternating I/O and computation of the same series (as long as there are more task series than cores).

The modes \texttt{fjio} and \texttt{fjc} consider the case of unbalanced amounts of I/O and computation tasks. \texttt{fjio} mode is particularly interesting as it allows to sustain more I/O requests per core when TASIO is in use. As long as the disk bandwidth is not saturated, running an I/O task with TASIO appears to be free because immediately after submitting the AIO requests, the runtime is able to run another task. The more I/O tasks that can be run sequentially in the same core, the more I/O petitions in flight the system will have a chance to optimize and process. When a computation task is encountered, the core is "stalled" and no more I/O requests can be issued from there until the task finishes. When the storage device is saturated, the TASIO effect is to only queue more I/O tasks and to run computation tasks earlier. However, when no more tasks are available, cores will idle until I/O petitions complete.

\subsection{Test Environment}
All tests have been run in a single node of Intel's Scalable System Framework (SSF) "Cobi" machine which features two Xeon E5-2690 v4 sockets with a total of 28 cores (56 hardware threads), 32KiB L1i and L1d caches, 256KiB L2 cache, 35840 KiB L3 cache, 128 GiB at 2400MHz of main memory, a 960 GB SSD Intel Optane 905P~\cite{intel:optane-ssd} used for the tests I/O operations and a SATA SSD which holds the system installation. The Linux kernel version is 4.10 and core frequency scaling is disabled system-wide.


\subsubsection{SSD Optane 905P Profiling}

We have profiled the Intel's 905P Optane SSD maximum random read and write speeds using the Flexible I/O tester (fio)~\cite{bench:fio}. The used fio configuration runs 56 threads (one per hardware thread) which issue up to four AIO requests of 4KiB and 1MiB. The results obtained in the 1MiB configuration closely resembles the official device specifications for sequential Read 2600 MB/s (2579.5 MiB/s) and sequential write (up to) 2200 MB/s (2098 MiB/s) as shown in table~\ref{tab:ssd-bench}.

\begin{table}[]
	\centering
	\begin{tabular}{|r|c|c|c|c|c|c|c|c|}
		\hline
		\textbf{Block Size} & \multicolumn{4}{c|}{\textbf{4KiB}}                & \multicolumn{4}{c|}{\textbf{1MiB}}                \\ \hline
		\textbf{AIO Depth}  & \textbf{1} & \textbf{2} & \textbf{3} & \textbf{4} & \textbf{1} & \textbf{2} & \textbf{3} & \textbf{4} \\ \hline
		\textbf{rand write} & 2255       & 2277       & 2285       & 2285       & 2278       & 2282       & 2283       & 2285       \\ \hline
		\textbf{rand read}  & 2265       & 2264       & 2264       & 2264       & 2548       & 2548       & 2547       & 2547       \\ \hline
	\end{tabular}
	\caption{Intel 905P Optane SSD throughput in MiB/s for block sizes of 4KiB (left) and 1MiB (right) and up to four (1,2,3,4) AIO petitions in flight}
	\label{tab:ssd-bench}
\end{table}

\subsection{Results}

We have run the TIOM benchmark for all combinations of computation time ranging from 1ms to 128ms with block sizes ranging from 4KiB to 8MiB in power of two steps. We have repeated this sequence for sequential read, sequential write, random read (rr), and random write (rw). Also, we have run the experiments using two configurations for the maximum number of tasks that can be run in parallel at the same time (this directly affect the number of task series as described in \ref{sec:tiom}). The configurations correspond to 128 and 256 which roughly corresponds to twice and four times the number of hardware threads respectively. Each configuration is repeated for the four TIOM operation modes \texttt{mix}, \texttt{1to1}, \texttt{fjio}, \texttt{fjc}, except for sequential read and write tests which were run only for the \texttt{mix} mode. Three versions of TIOM are benchmarked: a standalone version, a version preloaded with TASIO in blocking mode (bq) and a version linked with TASIO in non-blocking mode (nb). Each test finishes when a 20GiB file has been processed entirely (hence, the number of both I/O and computation tasks depends on the specified block size) and four repetitions are executed per configuration. However, we limited the execution time to 60s for each repetition.

Figures~\ref{fig:tiom-mix} and \ref{fig:tiom-fjio} show speedup for a selected subset of relevant \texttt{mix} and \texttt{fjio} configurations respectively. We are not showing the results for \texttt{fjc} and \texttt{1to1} because they did not prove to be relevant enough. In \texttt{fjc}, simulated computation is throttling too much I/O for TASIO to be effective, and in \texttt{1to1}, the results are quite similar to \texttt{mix}. Figure~\ref{fig:tiom-bw} shows bandwidth readings for both of the presented modes. The z-axis of all graphs either shows bandwidth (bw) readings in MiB/s for the standalone version or speedup (sp) readings in percentage achieved when comparing the standalone version with either the blocking (bq) or non-blocking (nb) versions. The left axis shows computation time in milliseconds and the right axis shows block size in KiB. White areas are close to 0\% speedup, green areas to positive speedup and red areas to slowdown. Bandwidth graphs have their own coloring scheme.

The bandwidth graphs of Figure~\ref{fig:tiom-bw} show that read operations are mostly able to saturate the disk consistently once the ratio $ \frac{block\_size}{comput\_time} \ge 64 $ is achieved. Write operations also reach the maximum speed rated by the manufacturer, but fail to keep it up as the block size increases.

As can be seen in both speedup Figures~\ref{fig:tiom-mix}~and~\ref{fig:tiom-fjio}, small blocks followed by long computation time lead to an underused storage device which is shown as a white triangle with its right angle pointing to the reader. As expected by the Amdahl's law, when computation far extends the I/O needs of an application, there is no point in using TASIO as the improvement is minimal.

Because read operations easily saturate the disk, we can only see the effect of using TASIO in the narrow and leaning diagonal region that drives the disk from underusage to saturation. Hence, TASIO non-blocking version is generally helping to saturate the disk in these cases. However, the blocking version sometimes leads to slowdown, quite likely because of the overhead introduced by creating and managing extra threads.

The difficulties presented to achieve sustained saturation throughput in write operations give TASIO the slack needed to actually improve the application performance. All write graphs show three common peculiarities: The first is, similarly to read operations, a diagonal of improvement that coincides with the device saturation ramp. The second is a moderate wavefront present after the diagonal for the biggest blocks which overlaps the throughput decrease seen in the bandwidth graphs. The third and last is a prominent peak standing at the smallest computation time values and between approximately 32 and 128 KiB that usually ranges between 40\% and 80\% but that it eventually reaches up to 100\% of speedup in cases such as figure~\ref{fig:mix_rw_128_spnb}.


Sequential I/O operations are slightly faster than random I/O operations. This leads to more room for TASIO to bring the device to saturation in the random case. In consequence, TASIO is, in general, able to achieve more performance (around 5\% increase) on the random I/O case. However, we are not showing the sequential I/O graphs because of their similarity with the random case. It is worth noting that a few sequential read tests reported slowdown around 15\%.

Although both TASIO modes achieve considerable speedups, the non-blocking mode is generally more efficient than the blocking mode. For instance, compare \ref{fig:mix_rw_128_spbq} with \ref{fig:mix_rw_128_spnb} or \ref{fig:mix_rr_128_spbq} with \ref{fig:mix_rr_128_spnb}. This is particularly true when a high degree of parallelism is present (256 parallel I/O tasks) and enough outstanding I/O requests are available as seen when comparing the \texttt{fjio} tests \ref{fig:fjio_rw_256_spbq} with \ref{fig:fjio_rw_256_spnb}. This makes sense as the higher the number of parallel tasks, the bigger the number of threads that are needed to processes more I/O operations in-flight, which leads to more overhead to manage them. Instead, no extra threads are required in non-blocking mode because tasks are not paused inside a thread context (blocking the entire thread), but tasks are detached of threads and remain in a "zombie" state until its associated pending events finishes.

Because write operations are more interesting than read operations (in this particular environment) from the TASIO point of view, Figures~\ref{fig:tiom-mix} and \ref{fig:tiom-fjio} only show random write tests when using 256 parallel tasks series. Read tests generally reported either minor speedup or slowdown when increasing parallelism. More parallelism means more overhead and because read improvements are limited, overhead exceeds the margin for improvement.

As mentioned before, the results obtained in both \texttt{1to1} and \texttt{mix} modes are quite similar. In practice, \texttt{1to1} tasks might be executed in a similar sequence as to how they would have been in \texttt{mix} mode, so there is not much difference appreciated. The \texttt{fjio} results shown in Figure~\ref{fig:tiom-fjio} achieve the highest speedups for specific write cases as seen in \ref{fig:fjio_rw_256_spnb}, but also show consistent slowdown regions. In general, this mode performs poorly on read operations such as in \ref{fig:fjio_rr_128_spbq}, hitting slowdown mostly in blocking mode but also in the non-blocking one.







\def \iogw {0.48\textwidth}

\begin{figure}[htb]
\centerline{
\subfloat[bw mix rw 128]{
	\includegraphics[width=\iogw]{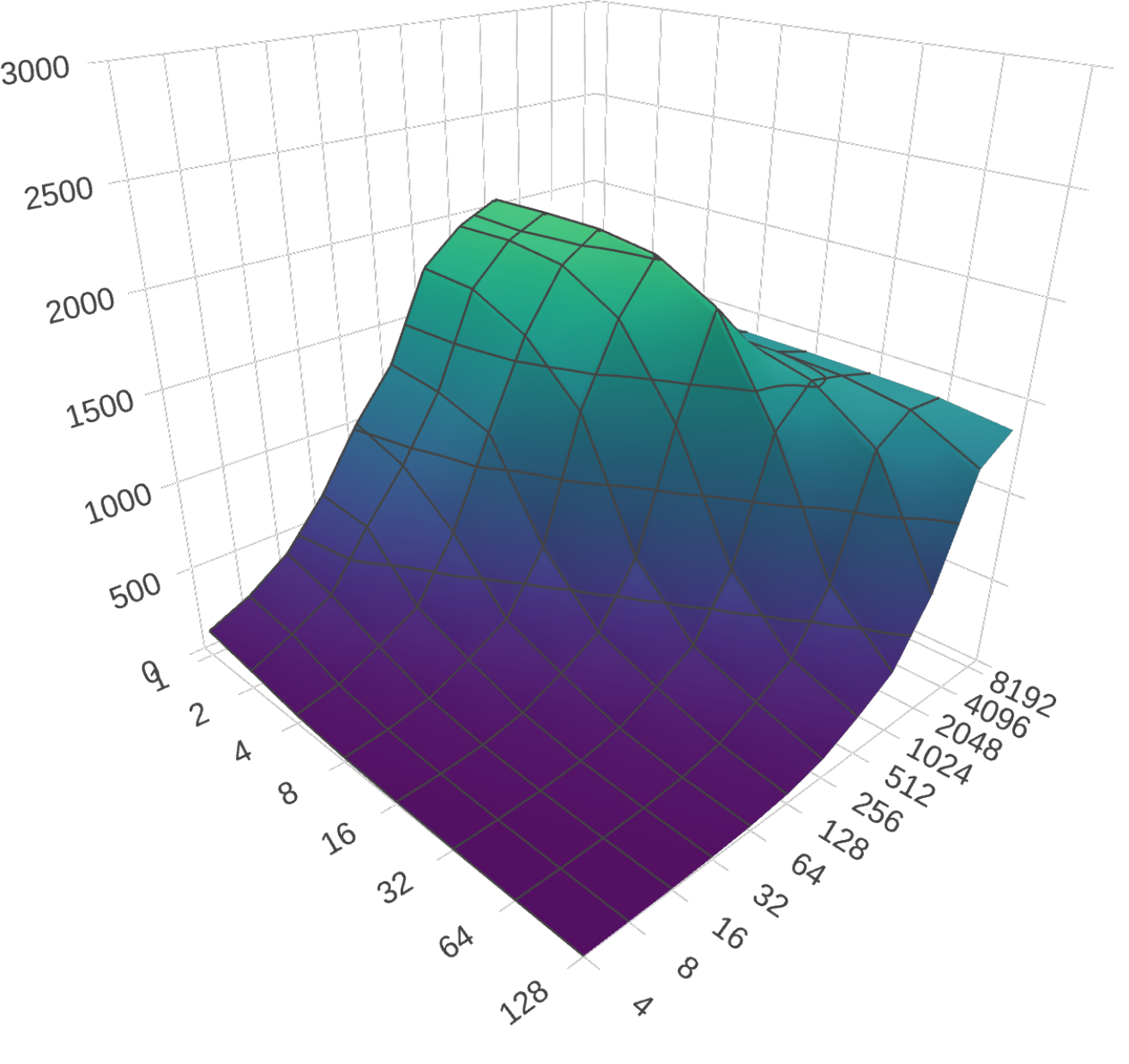}
}
\subfloat[bw mix rr 128]{
	\includegraphics[width=\iogw]{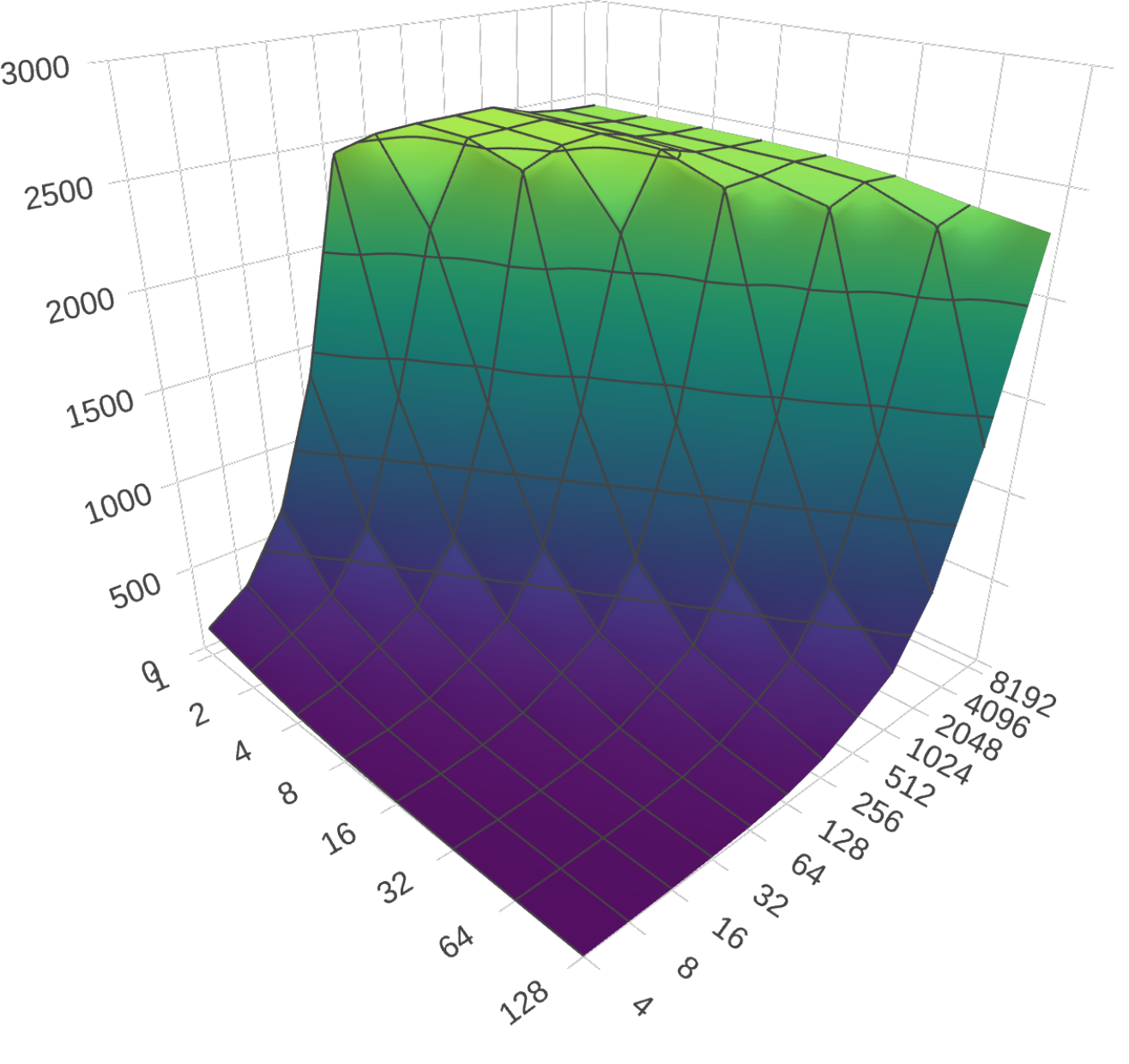}
}
}
\centerline{
\subfloat[bw fjio rw 128]{
	\includegraphics[width=\iogw]{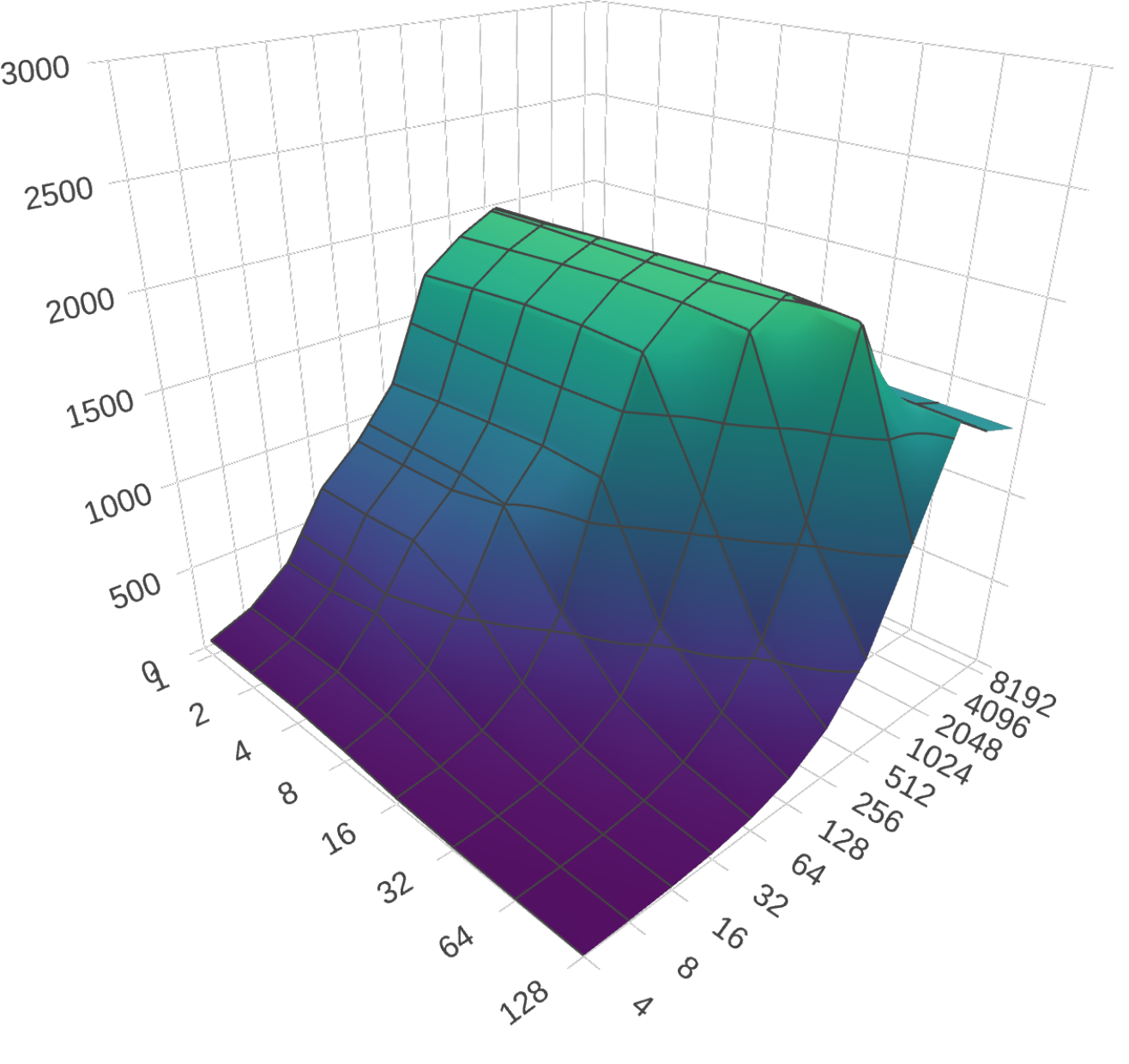}
}
\subfloat[bw fjio rr 128]{
	\includegraphics[width=\iogw]{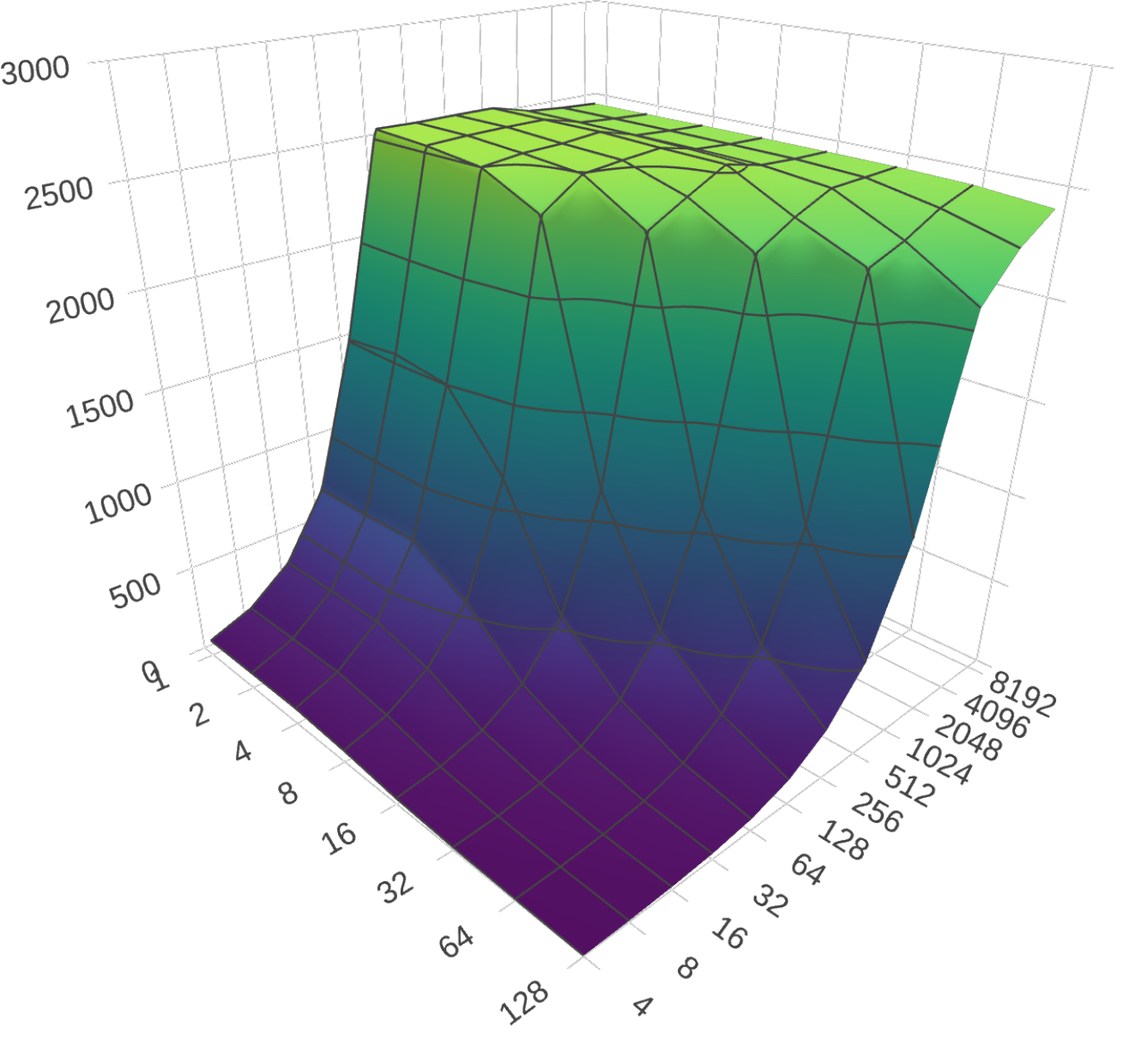}
}
}
\caption{TIOM storage I/O bandwidth tests for the standalone version. The upper axis shows bandwidth in MiB/s, the left axis shows simulated computational time in ms and the right axis shows block size in KiB.}
\label{fig:tiom-bw}
\end{figure}

\begin{figure}[htb]

\centerline{
\subfloat[sp rw bq 128]{
	\includegraphics[width=\iogw]{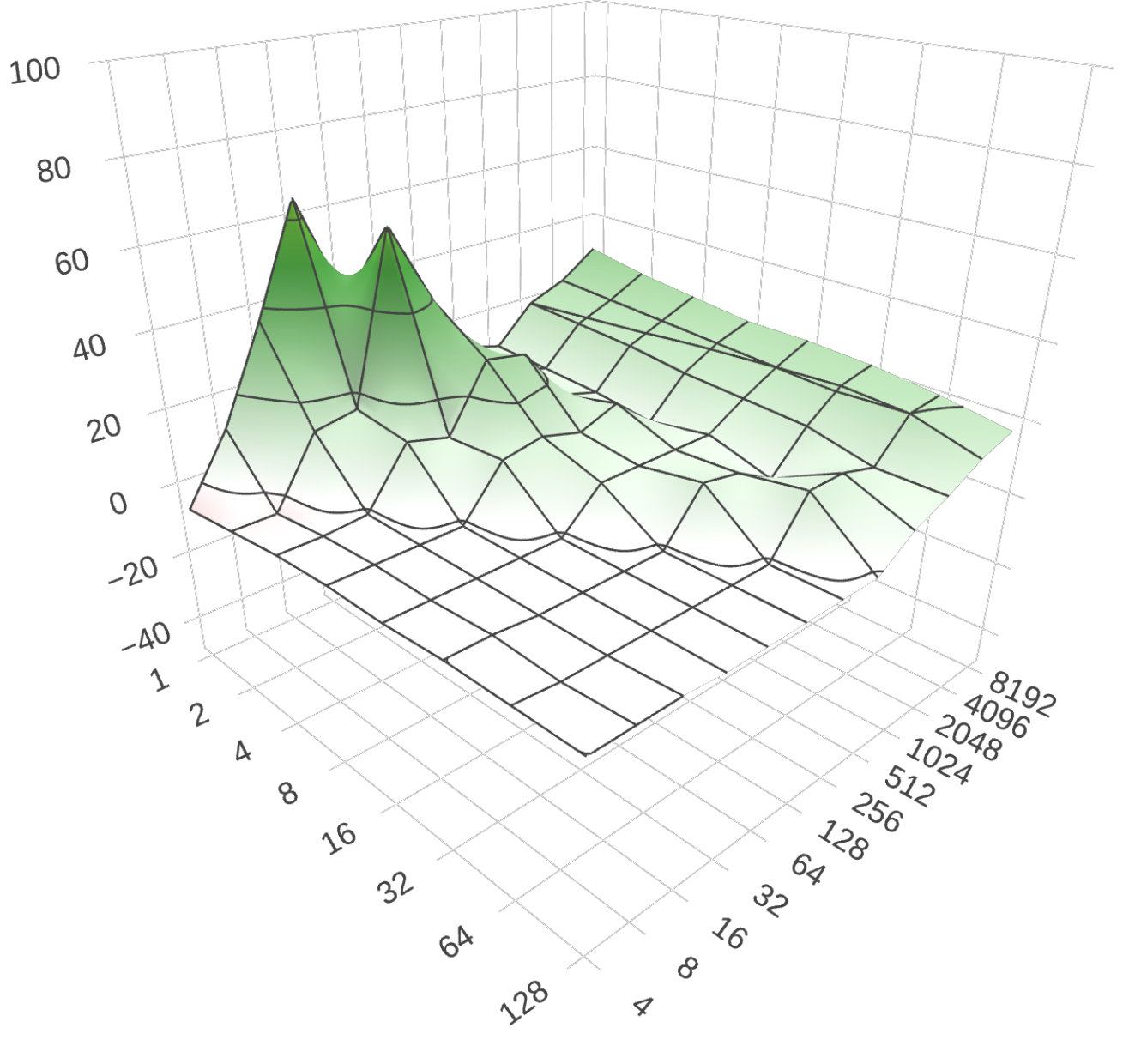}
	\label{fig:mix_rw_128_spbq}
}
\subfloat[sp rr bq 128]{
	\includegraphics[width=\iogw]{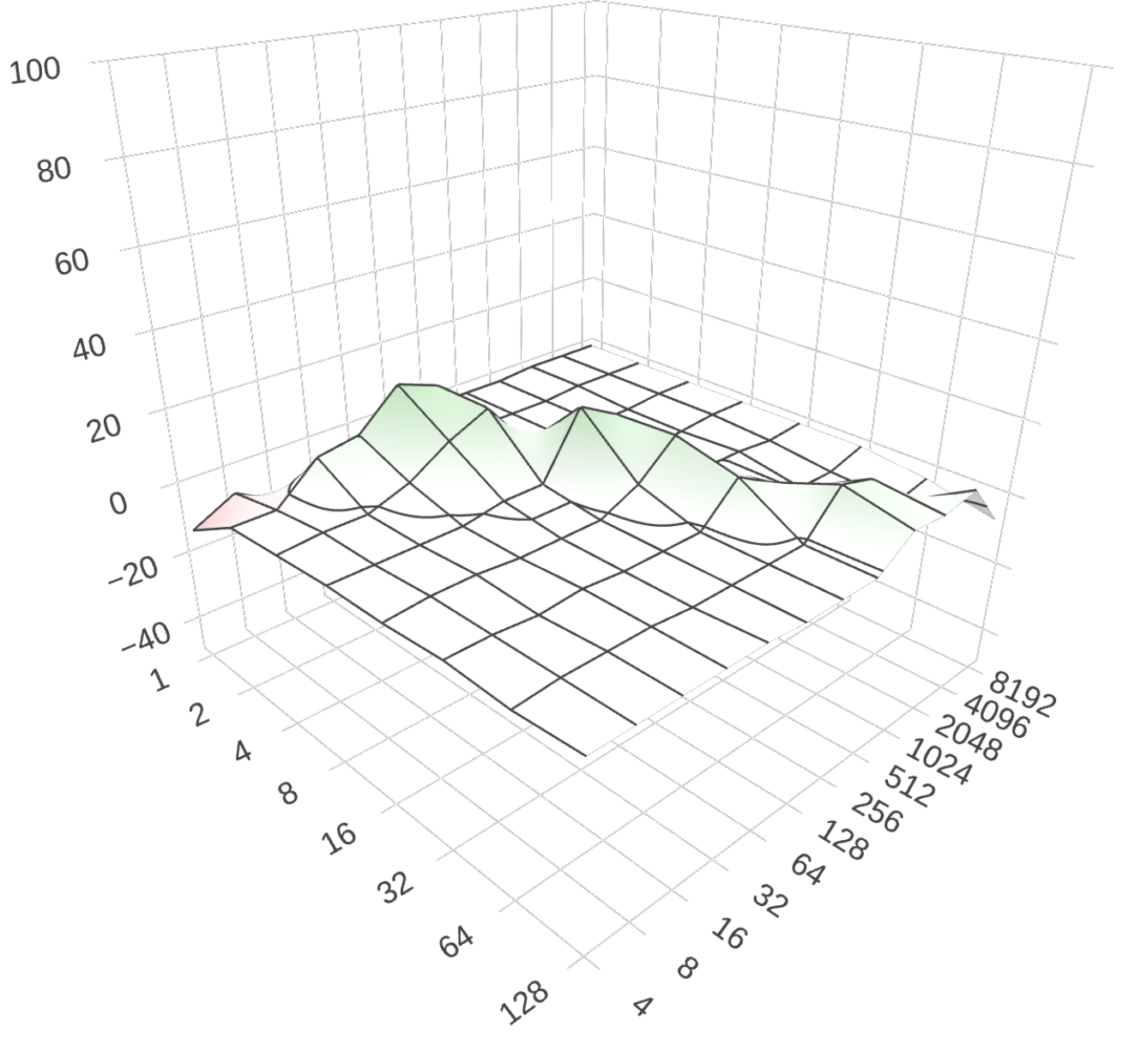}
	\label{fig:mix_rr_128_spbq}
}
}

\centerline{
\subfloat[sp rw nb 128]{
	\includegraphics[width=\iogw]{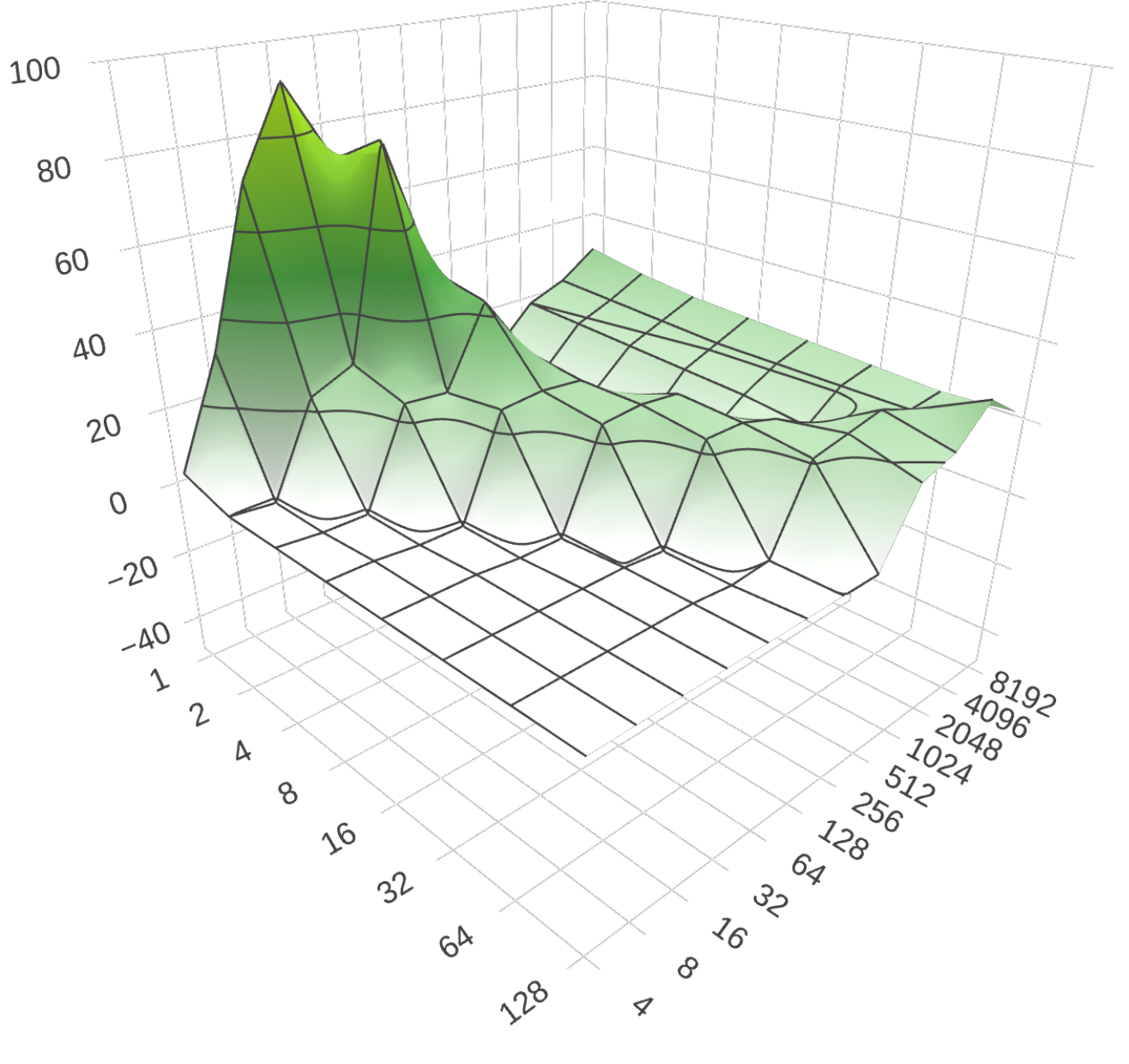}
	\label{fig:mix_rw_128_spnb}
}
\subfloat[sp rr nb 128]{
	\includegraphics[width=\iogw]{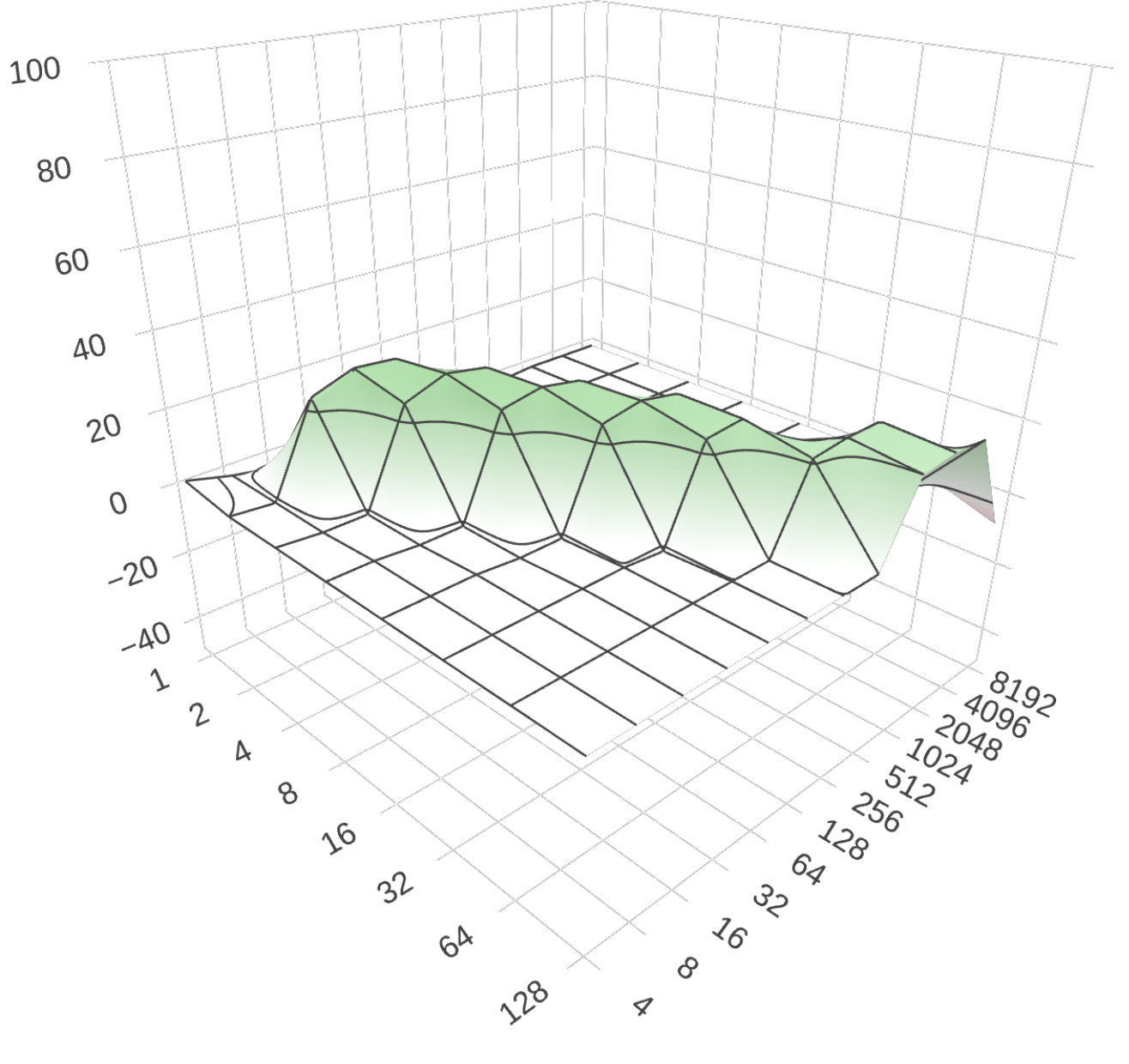}
	\label{fig:mix_rr_128_spnb}
}
}

\centerline{
\subfloat[sp rw bq 256]{
	\includegraphics[width=\iogw]{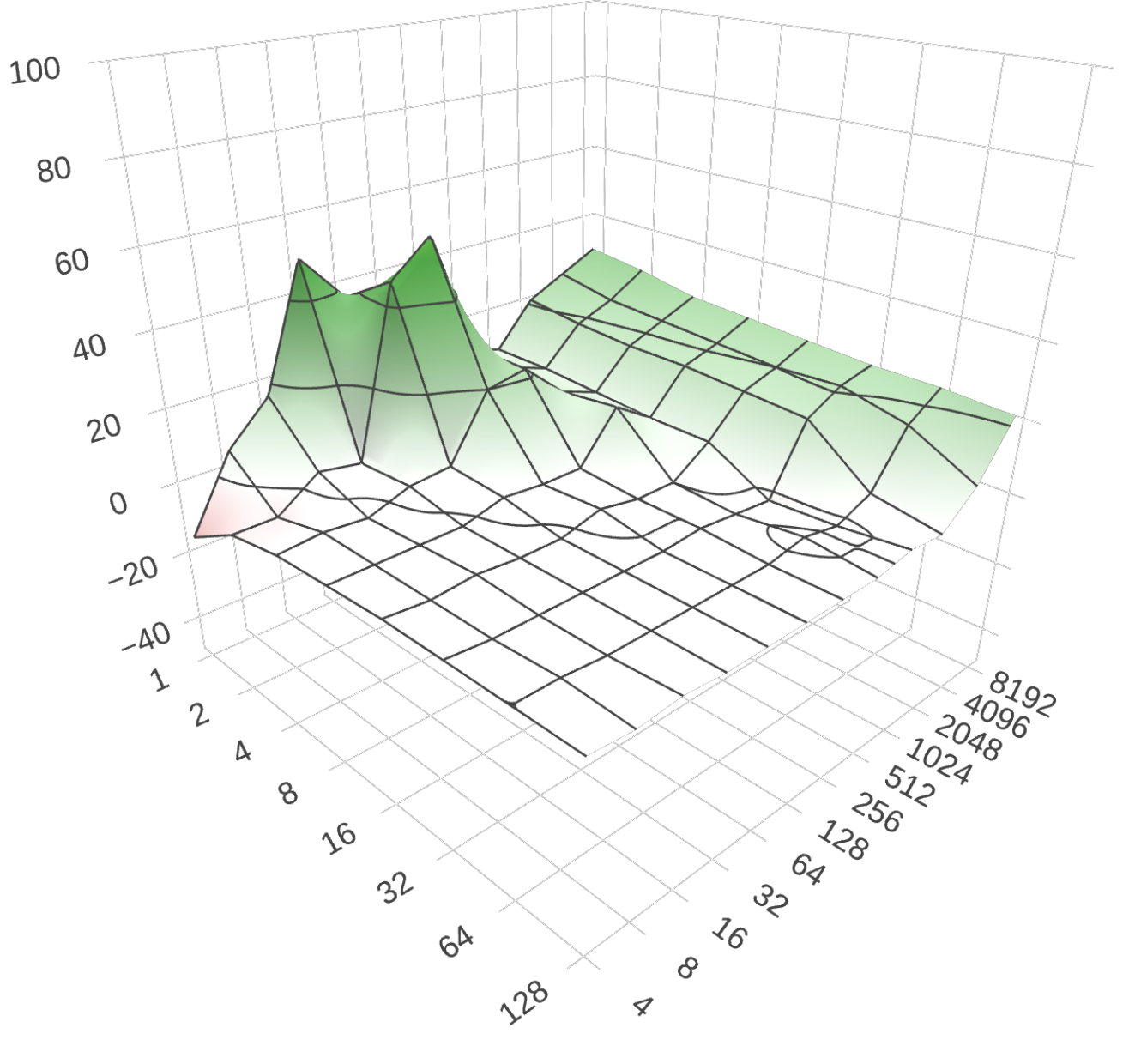}
}
\subfloat[sp rw nb 256]{
	\includegraphics[width=\iogw]{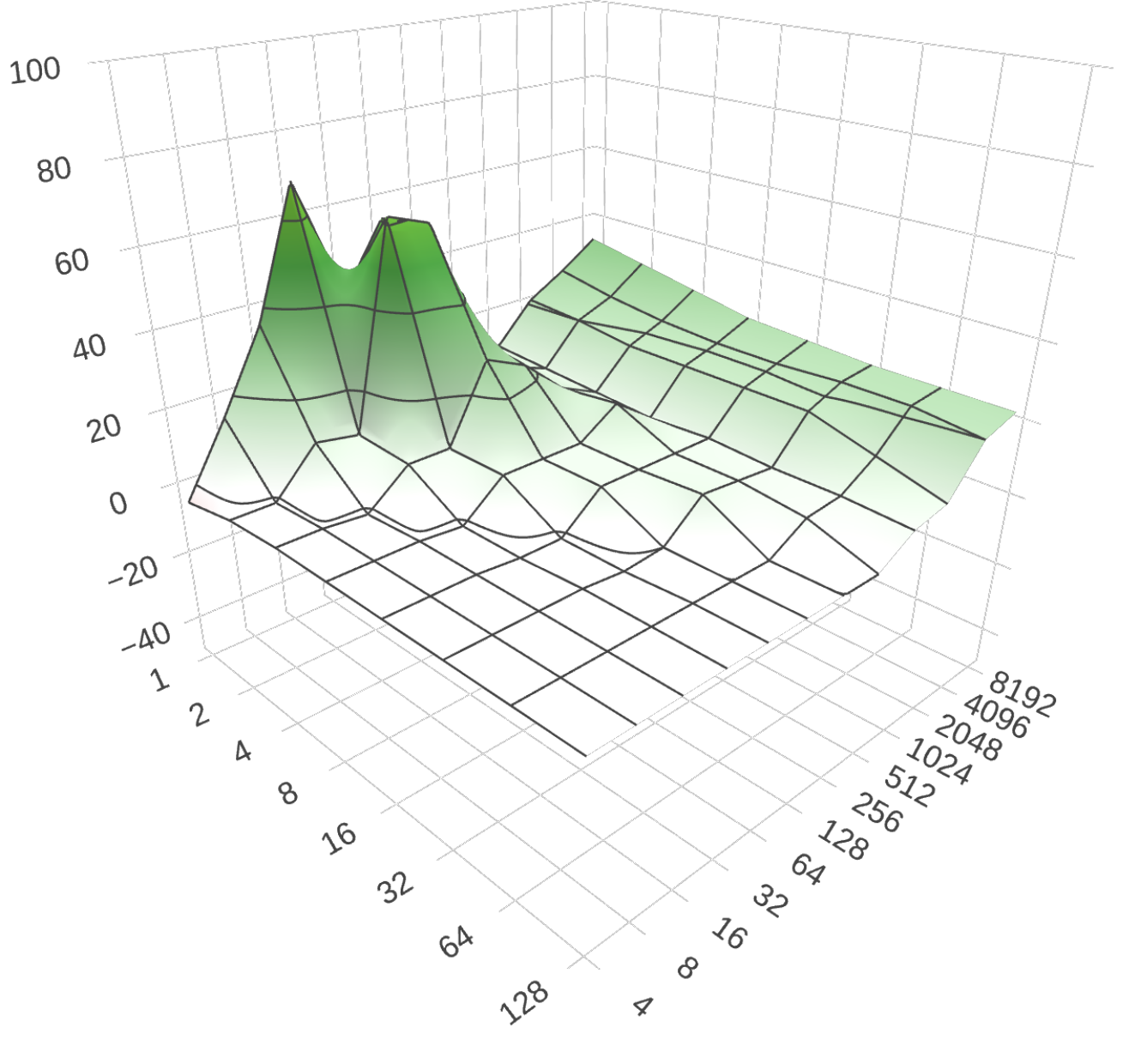}
	\label{fig:mix_rw_256_spnb}
}
}

\caption{TIOM tests for mix mode. The upper axis shows speedup, the left axis shows simulated computational time in ms and the right axis shows block size in KiB.}
\label{fig:tiom-mix}
\end{figure}

\begin{figure}[!htb]

\centerline{
\subfloat[sp bq fjio rw 128]{
	\includegraphics[width=\iogw]{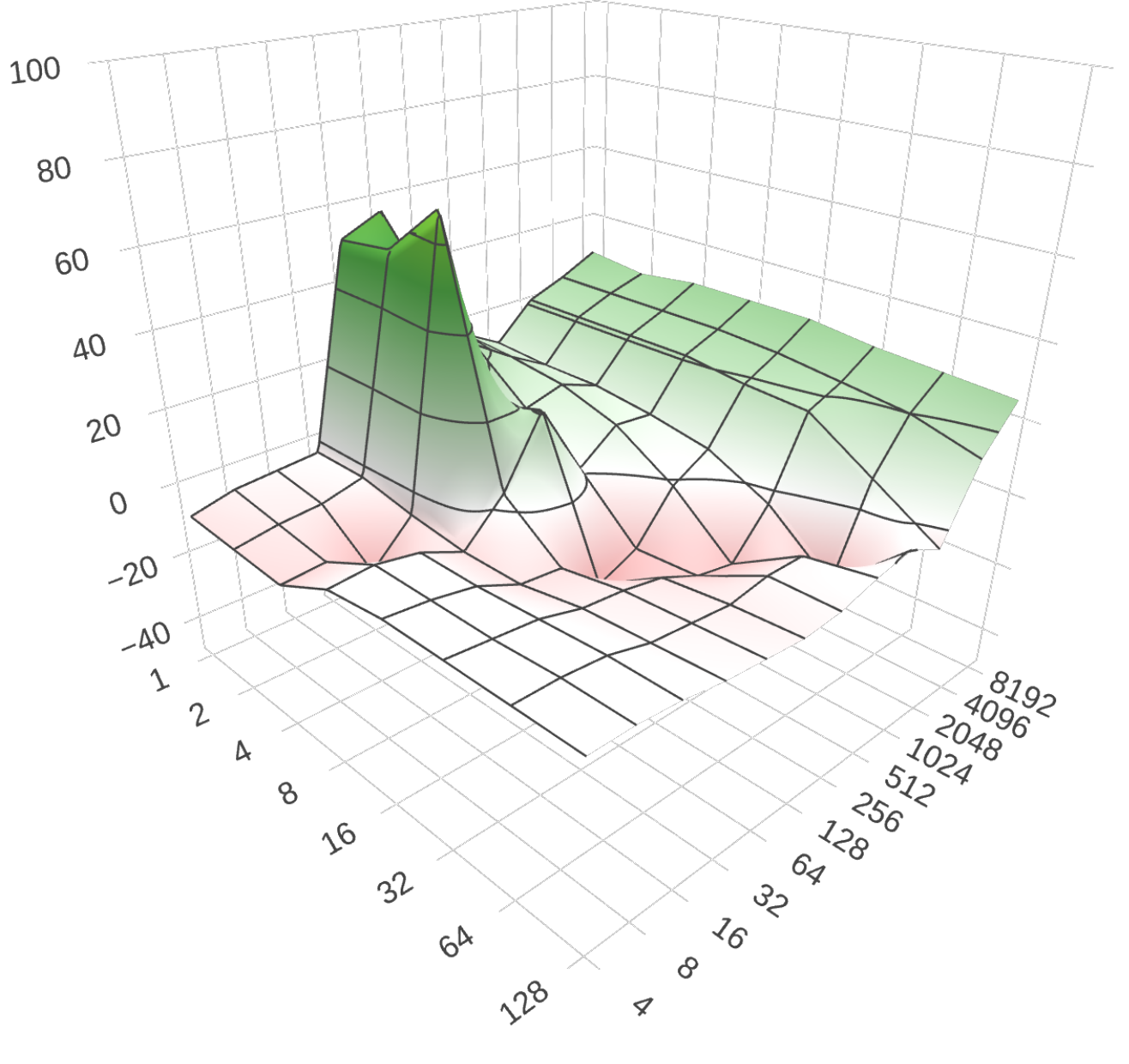}
}
\subfloat[sp bq fjio rr 128]{
	\includegraphics[width=\iogw]{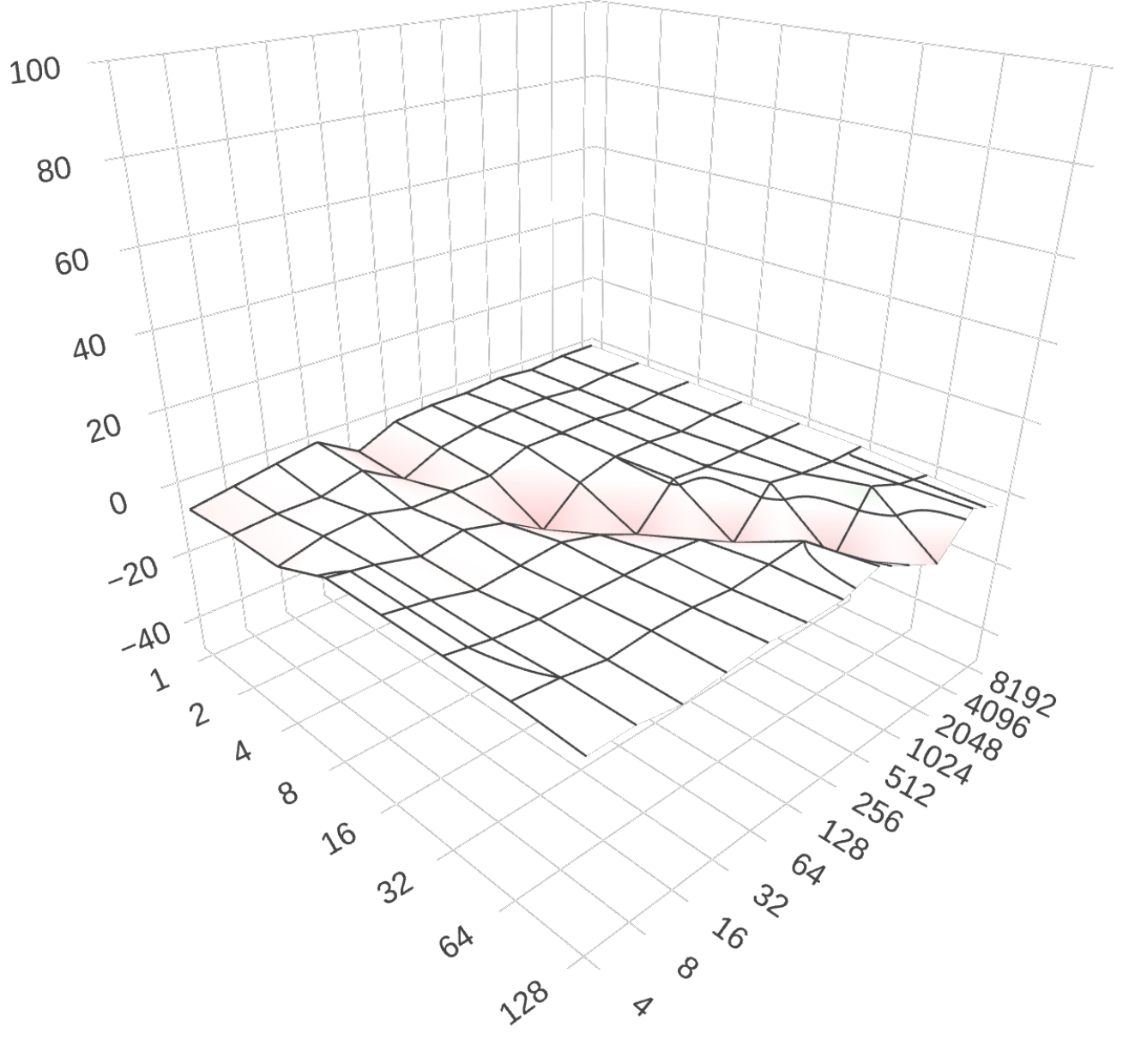}
	\label{fig:fjio_rr_128_spbq}
}
}

\centerline{
\subfloat[sp nb fjio rw 128]{
	\includegraphics[width=\iogw]{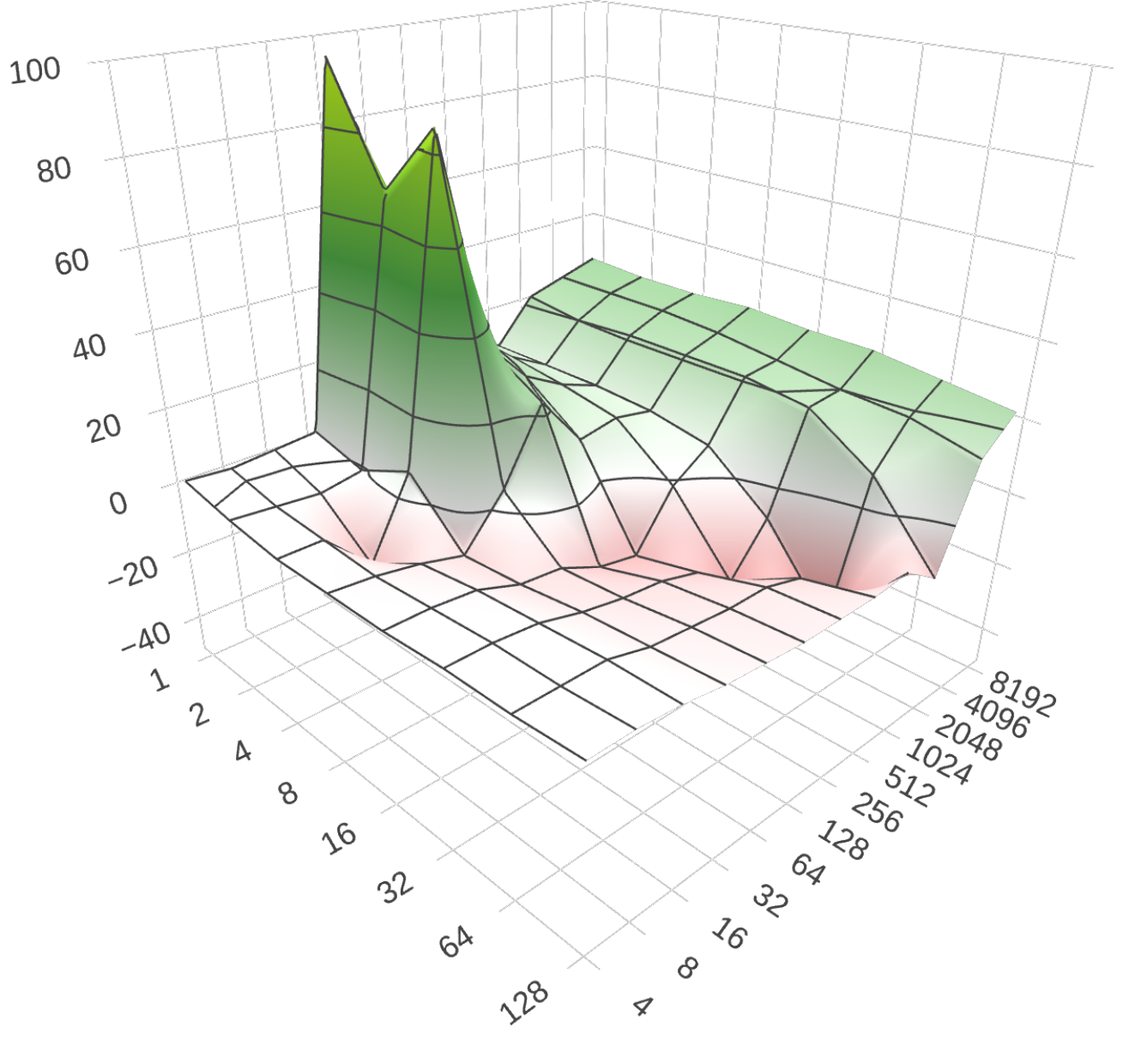}
}
\subfloat[sp nb fjio rr 128]{
	\includegraphics[width=\iogw]{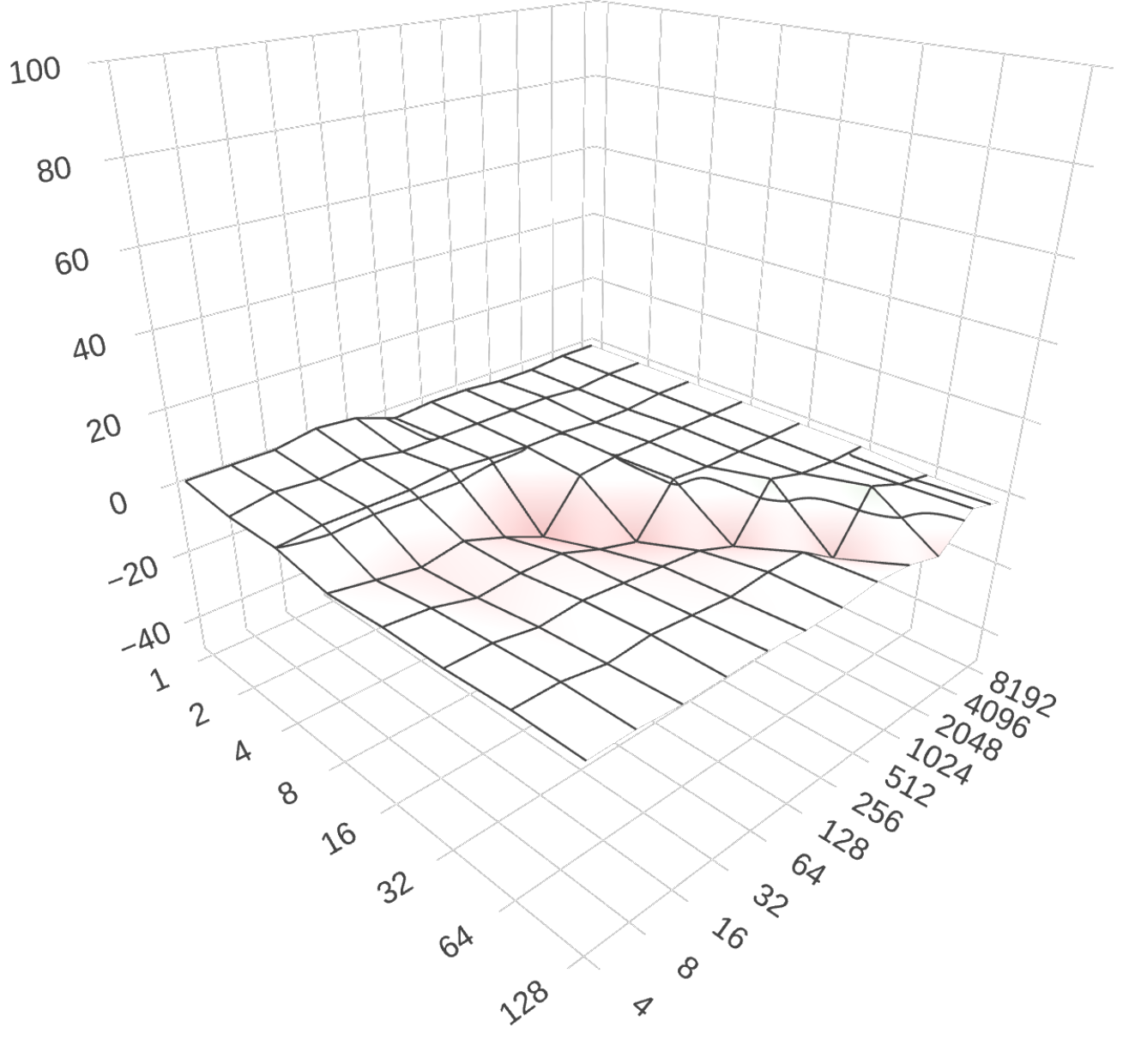}
}
}

\centerline{
\subfloat[sp bq fjio rw 256]{
	\includegraphics[width=\iogw]{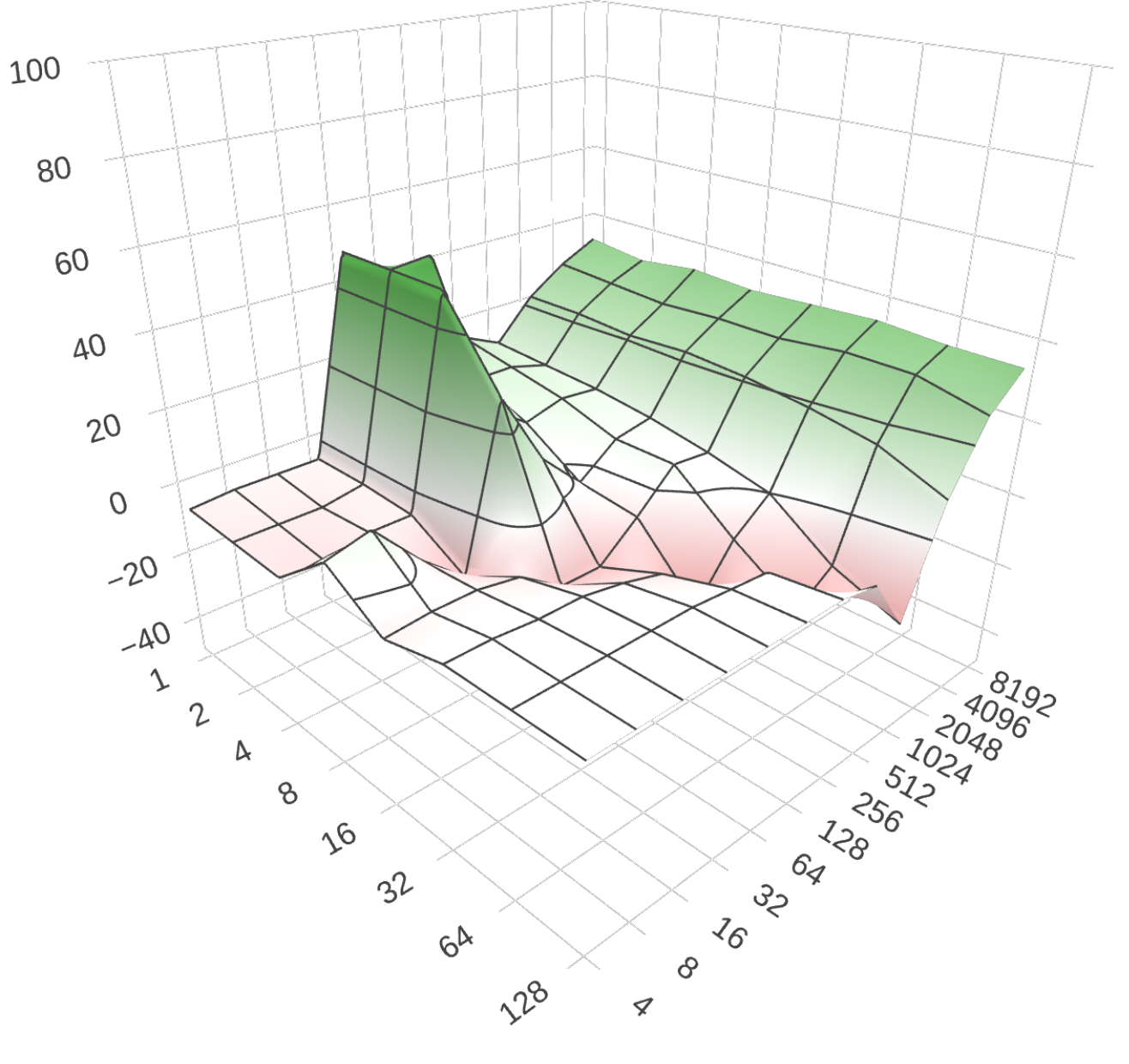}
	\label{fig:fjio_rw_256_spbq}
}
\subfloat[sp nb fjio rw 256]{
	\includegraphics[width=\iogw]{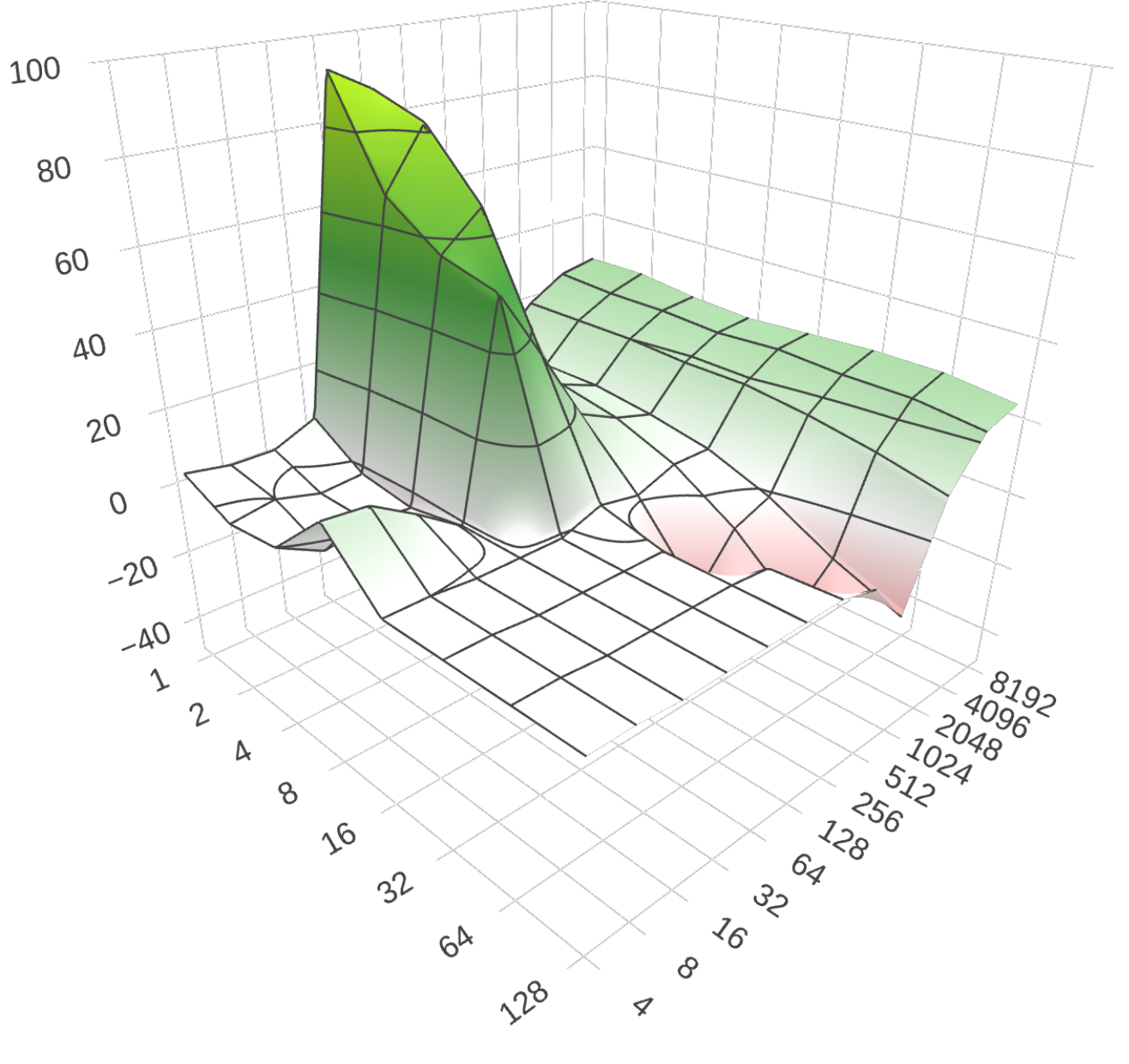}
	\label{fig:fjio_rw_256_spnb}
}
}

\caption{TIOM tests for fjio mode. The upper axis shows speedup, the left axis shows simulated computational time in ms and the right axis shows block size in KiB.}
\label{fig:tiom-fjio}
\end{figure}

\section{Conclusions And Future Work}
\label{sec:conclusions}

In this work, we have explored the state of the art of techniques to integrate storage I/O with the tasking model. We have presented the TASIO library to exploit such techniques in the context of read and write system calls and we have done exhaustive testing using a custom benchmark.

Both the blocking and non-blocking TASIO APIs have proved to improve the performance of the benchmark in most cases, although the blocking version's performance suffers due to the extra thread management overhead. Hence, the use of the library is encouraged but a previous analysis is needed to determine whether the application characteristics meet both TASIO and the system's AIO requirements which, in summary, are: 1) the application uses the disk intensively but 2) it is not already saturating it and 3) it does not benefit from the system's page cache and 4) I/O operation's meet the alignment and length requirements imposed by direct I/O and, finally, 5) there is enough computation work to be overlapped with I/O operations.

The exact parameters that fully exploit the library benefits are highly dependent on the system, with a primary focus on the number of cores, the storage device throughput and its capacity to sustain multiple parallel I/O requests. But for the particular set case tested in this work, we have found out that TASIO is able to achieve performance improvements between 40\% and 80\% (with peaks of up to 100\%) for write operations of around 32KiB to 128KiB interleaved with computation blocks of 1ms, but also speedups between 10\% to 40\% for block sizes greater than 1MiB run along computation tasks of any of the tested durations. The benefits of read operations are more discrete and its scope is limited to the narrow transition that leads to disk saturation, but still, a 20\% speedup is easily achievable when moving in this ranges. However, the question remains of which points that have been explored are really relevant to real applications.

Regarding our future work, we intend to test TASIO with real applications and to study the combined effect with the TAMPI library. We also plan to test an extra-thread-free TASIO blocking version. 

\section*{Acknowledgment}

This project is supported by the European Union's Horizon 2021 research and innovation programme under the grant agreement No 754304 (DEEP-EST), the Ministry of Economy of Spain through the Severo Ochoa Center of Excellence Program (SEV-2015-0493), by the Spanish Ministry of Science and Innovation (contract TIN2015-65316-P) and by the Generalitat de Catalunya (2017-SGR-1481). Also, the authors would like to acknowledge that the test environment (Cobi) was ceded by Intel Corporation in the frame of the BSC - Intel collaboration.

\clearpage

\bibliographystyle{splncs04}
\bibliography{ms}

\end{document}